\documentstyle{article}
\begin{document}
\textwidth=18cm
\textheight=20cm
\newcommand{\be}{\begin{equation}}
\newcommand{\ee}{\end{equation}}
\newcommand{\bc}{\begin{center}}
\newcommand{\ec}{\end{center}}
\newcommand{\fr}{\frac}
\bc
{\Large\bf
Viscous Big Bang Cosmology With Variable $G$ and  $\Lambda$}\\
\ec
\vspace{2cm}
\centerline{\Large Arbab.I.Arbab\footnote{E-mail: arbab@ictp.trieste.it
\& arbab64@hotmail.com}}
\bc
The abuds salam International Cenre for Theoretical Physics, ICTP, Trieste, 34100,
P.O. Box 586, ITALY.
\ec
\vspace{3cm}
\bc
ABSTARCT
\ec
{\large\sf We have examined some cosmological solutions with variable gravitational
and cosmological parameters and bulk viscosity. It is found that these solutions
are free from the cosmological problems of the standard model. The proposed
cosmology consists of two arbitrary parameters which are well constrained.
\vspace{2cm}
\\
{\it Heading: Freidman Models, Variable $G$ and $\Lambda$ models, Inflation and Bulk Viscosity} \\

Last Page  6
\newpage
{\bf Introduction}\\

The effects of viscosity in cosmology has been investigated by several authors
[1,2,3]. The bulk viscosity associated with grand unified theory phase transition
can lead to inflationary scenario [6].
\\

In spite of the success and internal consistency of the general theory of
relativity (GR), gravity is still on enigma at the microscopic level. In 1961 Brans
and Dicke offered an interesting alternative to GR employing Mach's principle [8].
However, cosmologically
the theory can still offer differences from GR. In particular, it can lead to $\dot G/G\sim H$, at 
the present epoch.
It also differs from GR in the early universe phase. It is shown by Canuto and Narlikar
that the $G$-varying cosmology is consistent with whatever cosmological observations
presently available [5]. In 1937 Dirac [12] proposed a model of a universe in which $G\propto t^{-1}$,
but his model ran in some serious difficulties. Recently, Abdel Rahman [7] and Massa [13]
proposed cosmological models in which $G$ is an increasing function of time.
\\

The variation of the gravitational parameter affects very much the Earth-Moon-Sun system.
From Newton's law and angular momentum conservation for the circular motion,
the orbital velocity $v$, the radial distance $r$ and the period $T$ of
revolution are given by: $v\propto G, r\propto G^{-1}$ and $T\propto G^{-2}$.
Therefore, an increase in $G$ would lead to a shortening of the Moon - Earth distance and the
Earth-Sun distance with the cosmic expansion.
The luminosity ($L$) of a given star is a sensitive function of $G$.
It is found that $L\propto G^4$ for a very massive star.
The variation of $G$ also affects the Earth temperature in the past [14].
\\

In this paper we enlarge our solutions by introducing variable gravitational and cosmological 
parameters and bulk viscosity.\\
This relaxation alleviates some problems of the standard model.
In particular, the age, horizon, missing mass, monopole can be reconciled with
the  present observations.\\
A notably consequence is that the inflationary solution can arise with variable
$G$ and $\Lambda$. We also note that our solutions become singular solutions when
$t\rightarrow 0$ (i.e. big bang ).
Our results overlap with those results obtained recently by Lima and Maia [11]
with constant gravitational parameter in their asymptotic regime.\\

{\bf The Model}\\
In a flat Robertson Walker (RW) metric, Einstein's field equations with time
dependent cosmological and gravitational parameters and a perfect fluid
with energy momentum tensor
\be
T_{\mu\nu}=(\rho+p)u_\mu u_\nu-pg_{\mu\nu}\ \ ,
\ee
give
\be
3\fr{\ddot R}{R}=-4\pi G(3p+\rho)+\Lambda\ \ ,
\ee
and

\be
3H^2=8\pi G\rho+\Lambda\ .
\ee
The Bianchi identities yield
\be
3(p+\rho)\dot R=-(\fr{\dot G}{G}\rho+\dot \rho+\fr{\dot\Lambda}{8\pi G})R
\ee
If we now consider an imperfect fluid (the one with bulk viscosity alone) the
effect is to replace, in the above equations, $p $ by $p-3\eta H$, where $\eta$ and $H=\fr{\dot R}{R}$ are the bulk
viscosity and Hubble parameter, respectively.\\ 
In this paper we will consider the form $\eta=\eta_0\rho^n$, $n, \eta_0 \rm \ \  consts.$.
\\
The equation of state relates the pressure ($p$) and the energy density ($\rho$)
of the cosmic fluid by the equation
\be
p=(\gamma-1)\rho
\ee
where $\gamma=\rm constant$.
In the presence of an imperfect fluid eq.(4) takes the form
\be
3(p^*+\rho)\dot R=-(\fr{\dot G}{G}\rho+\dot \rho+\fr{\dot\Lambda}{8\pi G})R
\ee
where $p^*=p-3\eta H$.\\
or
\be
3(p-3\eta H+\rho)\dot R=-(\fr{\dot G}{G}\rho+\dot \rho+\fr{\dot\Lambda}{8\pi G})R
\ee

The usual energy conservation equation is given by
\be
\dot\rho+3(p+\rho)H=0\ .
\ee
Imposing the above equation on eq.(7) one gets
\be
9\eta H^2=(\fr{\dot G}{G}\rho+\fr{\dot\Lambda}{8\pi G})\ .
\ee
Equation (5) and (8) give
\be
\rho=AR^{-3\gamma}\ \ , A=\rm const.
\ee
We will now consider other solutions that were not discussed in [4,15].
\\

{\bf 1. Solution with $\Lambda=0$\\}

Equation (3) gives
\be
8\pi G\rho=3H^2\ .
\ee
Equations (11) and (10) in eq.(9)  yield
\be
H=\fr{\gamma(2n-1)}{3\eta_0A^{n-1}}R^{3\gamma(n-1)}\ ,
\ee
hence
\be
R\propto t^{1/3\gamma(1-n)},\ \ n\ne1\ ,
\ee
and
\be
G\propto t^{\fr{(2n-1)}{(1-n)}}\ , \ n\ne1\  .
\ee
It is clear that in this model the gravitational parameter is an ever increasing
function of time. The case $n=1/2$ is not allowed in this model as this would
imply an empty universe.
The deceleration parameter is given by $q=-\fr{\ddot R R}{\dot R^2}=2-3n$ and
$H_0t_0=1/3(1-n)$ (the subscript 0 denotes the present day quantities).\\
To resolve the age problem with the present data ($0.6\le H_0t_0\le 1.3$),
one obtains the constraint $-0.25\le q_0\le 0.5 $ .
Note that the usual relation $\Omega=2q$ is not satisfied for the present model.

The inflationary solution in this model arises when $n=1$, which gives
$H=\fr{\gamma}{3\eta_0}$.
Therefore, one can have inflationary solution even if $\Lambda=0$ at the
cost of allowing $G$ to vary and introducing a bulk viscosity. Note also that the
energy density does not remain constant but decreases exponentially with time.
From above, we see that the smaller the coefficient of bulk viscosity ($\eta_0$) the bigger
the inflation rate. Thus the horizon, monopole, flatness problems are resolved.
\\

{\bf 2. Solution with $\Lambda=3\beta H^2$ and $\eta=0$}
\\

In the Freese {\it et al}. model [10] $x=\rho_v/(\rho_v+\rho)=\rm const \ (0\le x\le 1)$
, where $\rho_v=\fr{\Lambda}{8\pi G}$ is the vacuum energy density.
This is equivalent to setting $\Lambda=3\beta H^2$ in eq.(3),
 where $\beta$ replaces $x$.

Equation (9) then gives
\be
(1-\beta)\fr{\dot G}{G}+2\beta\fr{\dot\Lambda}{\Lambda}=0\ .
\ee
Upon using eqs.(3) and (10) the above equation gives
\be
R=Dt^{2/3\gamma(1-\beta)}, \ \  \ D=\rm const.
\ee
hence
\be
G\propto t^{2\beta/(1-\beta)}, \ \ \Lambda\propto t^{-2}, \ \ Ht=\fr{2}{3\gamma(1-\beta)}\ ,
\ee
where $\beta\ne 1$.\\
The deceleration parameter and the density parameter are given by
$q=(1-3\beta)/2$ and $\Omega=1-\beta$, respectively.
One can also write
\be
\Omega=\fr{2}{3}+\fr{2}{3}q\ .
\ee
Again to reconcile the age problem with the present data ($0.6\le H_0t_0\le 1.3$)
we impose the constraint $0 \le \beta \le 0.5 $. This gives $-0.25\le q_0\le 0.5$.
It is evident from eq.(16) that $G$ is an increasing function of time.
When $\beta=0$, we recover the Einstein de Sitter model.
The inflationary solution in this model is obtained when $\beta=1$ and is
given by $R=\rm const.\exp(\sqrt{\fr{\Lambda}{3}}t)$, which is the usual
de Sitter solution.
\\
{\bf 3. Solution with bulk viscosity, constant $G$ and variable $\Lambda$}
\\

In this case we take the Chen and Wu [9] ansatz, i.e.,  $\Lambda=\fr{3\alpha}{R^2}$,
$\alpha=\rm const.$. Equations (3), (9) and (10) yield
\be
R=Dt^{1/(2-3\gamma n)}, \ D=\rm const.,\ n\ne 2/3\gamma\ .
\ee
In this model $H_0t_0=1/(2-3n)$ and $q=1-3n$.
For age parameter in the range $ 0.6\le t_0 H_0 \le 1.3 $ one has $ 0.17 \le n \le 0.42$ .
We remark that for a physically acceptable solution $\Lambda < 0 $.\\
The inflationary solution in this model is obtained in the usual way
if $n=\fr{2}{3\gamma}$. For the radiation dominated universe
$n=\fr{1}{2}$ and hence $\eta\propto \rho^{1/2}$. This behavior has been found
by many authors to give a structurally stable solution.
\\
{\bf Concluding remarks}
\\

We have discussed some cosmological solutions with variable
cosmological and gravitational parameters and bulk viscosity. These
solutions are simple and rely on the conspiracy among $G, \Lambda $ and $\eta$
to satisfy the usual energy momentum conservation. This approach,
though not covariant, is an extension of the Einstein's general theory of relativity.
All solutions we have obtained are Big Bang solutions.\\
The inflationary solutions are obtained for all models. The age, horizon and flatness
problems are better resolved.\\ 
Some solutions are satisfied with low energy density.
The variation of the gravitational parameter required by the theory is found to be of the order
of Hubble parameter. This can not be excluded by the present data. However, the variation
of the gravitational parameter has immediate effects on the evolution of the
Earth-Moon-Sun system. More data are necessary to prove or disprove the variation of $G$.
This cosmology contains only two parameters viz. $\beta$ and $n$, which are constrained
by the conditions $0.17\le n\le 0.42$ and $0\le \beta \le 0.5$.
These conditions give consistent results for the observed cosmological data.
It is therefore evident that the introduction of bulk viscosity enriches the
proposed cosmology.
\\
{\bf References}
\\
$[1]$. M.Novello and R. A. Araujo, {\it Phys. Rev.D22}, 260(1980)\\
$[2]$. O.Gron, {\it Astrophys. Space Sci.173},191(1990)  \\
$[3]$. J.D.Barrow, {\it Nucl. Phys. B310}, 743(1988)\\
$[4]$. A.I.Arbab, {\it Gen. Rel. Gravit. 29, 61 (1997)}\\
$[5]$. V.Canuto and V.J. Narlikar, {\it Ap.J 236(1980) } \\
$[6]$. G. Murphy, {\it Phys. Rev. D8}, 4231(1973)    \\
$[7]$. A-M.M.Abdel Rahman, {\it Gen. Rel. Gravit.22}, 655(1990)\\
$[8]$. C.Brans and R.H.Dicke, {\it Phy. Rev. D41}, 695(1961) \\
$[9]$. W.Chen and Y-S.Wu {\it Phys, Rev.D41},695,(1990)\\
$[10]$. K.Freese, F.C.Adams, J.A.Frieman and E. Mottola {\it Nuc.Phys.B287},
797(1987)\\
$[11].$ J.A.S.Lima and J.M.F.Maia, {\it Phys. Rev.D49}, 5597(1994)\\
$[12]$ P.A.M.Dirac, {\it Nature, 139, 321(1937)}\\
$[13]$ C.Mass, {\it Astrophys. Space Sci., 232,143(1994)}\\
$[14]$ T.L.Chow, {\it Nouvo Cimento Lett. 31, 119(1981)}\\
$[15]$ A.I.Arbab, {\it Nouvo Cimento {B113,403(1998)}\\
}

\end{document}